# Countering Social Engineering through Social Media: An Enterprise Security Perspective

Heidi Wilcox and Maumita Bhattacharya
School of Computing & Mathematics
Charles Sturt University, Australia
{hwilcox,mbhattacharya}@csu.edu.au

**Abstract.** The increasing threat of social engineers targeting social media channels to advance their attack effectiveness on company data has seen many organizations introducing initiatives to better understand these vulnerabilities. This paper examines concerns of social engineering through social media within the enterprise and explores countermeasures undertaken to stem ensuing risk. Also included is an analysis of existing social media security policies and guidelines within the public and private sectors.

## 1  Introduction & Background

Social media sites such as Facebook, Myspace, LinkedIn, and Twitter are a data mining goldmine for readily available personal and sensitive information made publicly for the web, especially when the majority of participants are using default privacy settings. (King, 2008; Furnell, 2008; Slonka, 2014; Nayak Prince & Robinson 2014; Wong et al 2014). The increased adoption of social media technologies and failing to protect company information may result in data leakage, business continuity failures and compliance breaches, reputational risks through loss of valuable intellectual property, consumer confidence and competitive advantage (Colwill, 2009; Almeida, 2012). Traditional security countermeasures are not keeping up with these changes in the workplace as more businesses are encountering breaches targeting the human elements, such as social engineering. (Colwill, 2009; Rudman, 2010; He, 2012). Social engineers exploit human behaviour idiosyncrasies to form an attack from the outside that leads them to gain inconspicuous entry into protected areas of the company for their own illicit use. (Mitnick & Simon, 2001).

As the line between business use and personal use is often blurred, social engineers can gather sensitive data from any number of social media accounts to form a personal resume on a targeted employee. (Meister & Willyerd, 2010). Traditionally, email was the primary vector for spam and phishing exploits, however, the popularity and scope for large volume targets in social media has seen these threats moving away from email and on to social platforms. Web based attacks such as phishing are consistently found to be the leading transport vectors for cyber-attacks; social media and social gaming provides the perfect vehicle or attack surface for delivering lures and payloads. (Arachchilage & Love, 2014; Ikhalia, 2014).  The top three social media issues negatively experienced by organizations include: employees sharing too much information, the loss of confidential information, and increased exposure to litigation (Symantec, 2011). Other equally important results include losses concerning employee productivity and increased risk of exposure to virus and malware (Almeida,

2012). These platforms enable social engineers to operate freely, efficiently and cost effectively with low margins for getting caught. (Franchi, Poggi & Tomaiuolo, 2014).

Boudreaux (2010a), and Foreshew (2012), propose that organizations protecting information assets through effective security policies and governance will more effectively manage the business risks of the future. Social media policies and guidelines provide advice on how social media participation will be applied to all of the members of an organization (Bell, 2010). It is also reported that the most effective security countermeasure against social engineering is to increase employee awareness of the many tricks employed by social engineers against them in the workplace. (Bada & Sasse, 2014).

The remainder of the paper is organized as follows. In Section 2 we explore those countermeasures currently offered by enterprise to address the challenges faced by social engineering through social media concerning people, process and technology. We also review existing social media policies from opposing sectors and compare areas of coverage in Section 3 with some concluding remarks presented in Section 4.

## 2 Countering Social Engineering through Social Media: Current Perspectives

The following global perspectives underline information security practices concerning people, process and technology that are currently used by enterprise in an attempt to decrease loss attributed to their social media usage.

### 2.1 People

Global organizations are embracing new technologies that explore huge business benefits but also bring catastrophic organizational risk. (Almeida, 2012). Countermeasures for online social engineering concerning people and employees have had various levels of success. Information security taskforces aligned to assess such threats are now heading in a positive direction towards understanding the motivations behind these attacks. These paths include establishing types of threat vectors and introducing awareness initiatives that effectively reduce business risk. (VMIA, 2010). Current practices focus on creating individual employee awareness and training in both the public and private sectors; whether they are online at home or work. There is a general consensus from both sectors that there needs to be collaboration from government and business in all industries to increase cyber security effectiveness.

The UK and US governments are guiding business and consumers with awareness initiatives in response to a trend in developed nations to adopt a web 2.0 and social business model framework for all government departments and processes. Recent awareness campaigns include the introduction of a national computer emergency response team (CERT-UK) in the UK to focus on new cybersecurity policies (Luxford, 2014) and a '10 steps to cyber security' awareness program; The Cyber Security Awareness Campaign organized by the National Institute of Electronics and Information Technology in India; Go Safe Online (including annual Awareness Day) in Singapore and national Cybersecurity Policy Review in 2013 from the US government initiated a 'Stop.Think.Connect' Awareness campaign with a national Awareness month.

Australian culture embraces mateship and social interaction as an integral part of everyday life. This intrinsic character trait therefore carries over to the workplace to form primarily trusting and open social norms between the organization's management, employees and customer base. By adopting social media as a technological tool to unite business processes, Australians – as employers or employees – are prime targets for social engineers to exploit this trusting 'weakness'. (Mitnick & Simon, 2001).

In response to the increasing threats to online security the Australian government under Prime Minister Julia Gillard introduced an amalgamation of cyber security and information security specialist organizations to form the Australian Cyber Security Centre (ACSC). The Centre will host experts in their fields representing Defense Signals Directorate, Defense Intelligence Organization, the Australian Security Intelligence Organization, the Attorney General's Department's CERT-Australia, Australian Federal Police and Australian Crime Commission. The initiative will analyze and assess incoming cyber threats, and work closely with various industries and private sector parties to formulate effective countermeasures and create public awareness. Other major Australian awareness initiatives already in place include Stay Smart Online including an annual Awareness Week, the annual Cyber Security Challenge Australia, and Cybersmart aimed at educating schools and school age children.

### 2.2  Policy

According to PriceWaterhouseCoopers (2013), fewer than 30% of global organizations have existing policies that include countermeasures covering social media usage. Results from this paper's policy review also found the vast majority of reviewed policies from global organizations did not include countering measures for dealing with online social engineering or phishing attempts while participating in social media platforms. General inconsistency of current policy design suggests that organizations are disparate in their views to countermeasure social media risk, and are lacking clarity in ways to precede legally and ethically.

In 2010, governments within the US and Australia declared the Government 2.0 initiative which would allow communications between Government and the people to be 'open' and transparent. (AG, 2010). In response a major restructure of current security practices ensued. These governments are still trying to figure out how to put boundaries around an employee's personal, professional, and official agency use. Each use has different security, legal, and managerial implications and government agencies are tasked with striking a balance between using social media for official agency interests only, and allowing all employees access for personal and professional interests.

The majority of Government agencies are managing online and social media access in two ways:

1. Controlling the number or types of employees who are allowed access to social media sites or

2. Limiting the types of sites that are approved for employee access.

A third approach sees agencies providing customized social media channels, such as GovLoop (2014), 'in-house' for employees, behind the organization's firewalls.

Effective examples of Australian guidelines can be viewed from The Australian Taxation Office (ATO), Department of Immigration and Citizenship (DIAC) and Department of Human Services (DHS). Additionally, The Australian Signals Directorate (ASD) in conjunction with The Department of Defense maintains a wealth of all access publications for all industries including the comprehensive Australian Government Information Security Manual which is produced by ASD to guide Australian Government ICT professionals on security of national systems.

Despite this proactive approach by government, private sectors seem to be looking cautiously inwards for security research and strategy. The Ponemon Institute (2011) observed that only 33% of private organizations in Australia had adequate policy relating to social media usage. Some of the better examples of private sector policy development include Telstra, Dell, IBM and Kodak. US Tech giant CISCO has also released a comprehensive policy duo with the CISCO Social Media Playbook: Best Practice Sharing; and the Social Media Policy, Guidelines & FAQs.

The authors of this paper also undertook an analysis of social media policies currently existing in social media management for organizations. Tables 2 and 3 have been drawn to compare current standards for 24 organizational social media policies and guidelines. These policies were selected randomly from the online publically available database of social media governance documents compiled by Chris Boudreaux (2014). A balance was sought between public and private sectors in policy selection (19 public and 19 private sector) for comparison. The samples were limited to those obtainable on publically based internet search and were further limited to those policies written in English (purposed for the understanding of the author). This small cross section acted as a guide to contribute to the author's understanding of the research topic. Delivering the information in table format provides a clarity to observing a pattern of social media issue coverage areas included as a general rule within these documents. The aim for these tables is to provide review of as much documentation of current, publicly available policy covering organizational use of social media to formulate resources for a best practice framework for future research direction. Primary to our purpose of analysis, these policies were gathered to investigate coverage of security risks to employees and advice on mitigating social engineering through these technologies. To aid in achievement of this goal the documents were coded with the use of Atlas.ti software. The qualitative process involved coding of key points from all 24 documents with Atlas.ti (2015) which revealed groupings of the data set for comparison. The comparison through Atlas.ti coding and extraction ascertained a trend for six primary coverage areas (Table 1) most likely to appear in a policy or guideline to act as educational advice to employees, or as a means of legal protection for the organization. As the documents varied widely in their approach on an individual level, this representation will serve as a general overview of issues addressed and listed for employee guidance. This is certainly the case when dealing with new technologies such as social media adoption. Social engineers focus on people as their targets primarily, using the human element to cause corruption to their own organizational technical tools and assets to infiltrate deeper within the organization

Table 1. Description of the six social media coverage areas used for policy review.

| Coverage Area & Description | |
|---|---|
| *1 Social Media Account Management* | *Policy Coverage Percentage* |
| Applies to the creation, maintenance and destruction of accounts designed for organizational purposes. Official accounts are predominantly operated by designated employees trained or certified in social media communications security. | N=11 46% |
| *2 Acceptable Use* | *Policy Coverage Percentage* |
| This area of policy is relevant to how a company's employees will use social media technologies either at work in a professional capacity, or for personal use while at work. Policies may state if certain sites are restricted from use, and also if personal use is condoned while on working time. Consequences of any violations would be listed here. | N=23 96% |
| *3 Social Media Content Management* | *Policy Coverage Percentage* |
| In conjunction with account access and creation, policies need to cover what content is to be published online. Official accounts usually have responses drafted from CEO's or Department Managers, carefully sanctioned by organizational leaders. Professional accounts serve the purpose for regular business processes such as customer support or marketing (similar to email functionality). | N=15 62% |
| *4 Employee Conduct* | *Policy Coverage Percentage* |
| Guides and policies include terms of ethical use and conduct for those participating in work-related online engagement. Most policies will either refer to existing Codes of Conduct or elaborate further into the "do's and don'ts" of social media use. | N=21 87% |
| *5 Legal* | *Policy Coverage Percentage* |
| Policy coverage includes any reference to laws and regulations that may have an impact on the company or the individual while using online communications. This can be a general point where employees are expected to abide by applicable laws but with no mention of any one in particular. Some policies elaborate further by explaining impact to action with each specific law, especially privacy and confidentiality and; copyright and intellectual property. | N=22 92% |
| *6 Security* | *Policy Coverage Percentage* |
| Policies cover technical and behavioural security issues. Of particular interest to this research is the inclusion of advice or awareness to the risks attributed to social engineering. It is imperative to include awareness on social engineering attacks (spearfishing, click-jacking) directed at employees using social media information. Technical security measures such as password protection, identification authentication (PKI), and virus scans can be included here. | N=10 42% |

Table 2. Social media policies reviewed from the public sector.

| Social Media Policies- Public Sector | | | | | |
|---|---|---|---|---|---|
| **Organisation** Department of Information Technology Government of India | | | | | |
| **Policy Type** Social Media Guide | | | | | |
| **Objectives** Provides a basic framework and guidelines for government agencies and e-projects | | | | | |
| Social Media Account Mgt$_1$ | Acceptable Use$_2$ | Social Media Content Mgt$_3$ | Employee Conduct$_4$ | Legal$_5$ | Security$_6$ |
| ✓ | ✓ | ✓ | | ✓ | ✓ |
| **Organisation** US Federal CIO Council | | | | | |
| **Policy Type** Social Media Guide | | | | | |
| **Objectives** Provide best practices and recommendations for federal use of social media and cloud computing resources | | | | | |
| | ✓ | | | ✓ | ✓ |

| Organisation | | | | | | |
|---|---|---|---|---|---|---|
| **Organisation** US Environmental Protection Agency | | | | | | |
| **Policy Type** Social Media Policy | | | | | | |
| **Objectives** This policy applies to EPA employees, contractors, and other personnel acting in an official capacity on behalf of EPA. | | | | | | |
| | ✓ | ✓ | ✓ | ✓ | | |
| **Organisation** Harvard University | | | | | | |
| **Policy Type** Social Media Guide | | | | | | |
| **Objectives** For those required to speak on behalf of Harvard at an individual level. | | | | | | |
| | ✓ | | ✓ | ✓ | | |
| **Organisation** US Department of Defence | | | | | | |
| **Policy Type** Social Media Policy | | | | | | |
| **Objectives** For all members of all departments in US defence force while using internet services including social networking and twitter | | | | | | |
| | ✓ | ✓ | | ✓ | | |
| **Organisation** UK Ministry of Defence | | | | | | |
| **Policy Type** Social Media Policy | | | | | | |
| **Objectives** Enables Service and MOD personnel to make full use of online presences while protecting their own, Service, and Departmental interests | | | | | | |
| ✓ | ✓ | ✓ | ✓ | ✓ | ✓ | |
| **Organisation** National Library of Australia | | | | | | |
| **Policy Type** Social Media Policy | | | | | | |
| **Objectives** Applies to all Library employees using or having a need to participate in online social media activity for official Library communications and through personal accounts which they have created and administer themselves. | | | | | | |
| | ✓ | ✓ | ✓ | ✓ | ✓ | |
| **Organisation** Department of Justice Victoria, Australia | | | | | | |
| **Policy Type** Social Media Policy | | | | | | |
| **Objectives** Recommended for members and contractors of the department for online use | | | | | | |
| ✓ | ✓ | ✓ | ✓ | ✓ | | |
| **Organisation** Australian Public Service Commission | | | | | | |
| **Policy Type** Social Media Guide | | | | | | |
| **Objectives** Guidance for APS employees and Agency Heads to help APS employees understand the issues to take into account when considering making public comment, including online | | | | | | |
| | ✓ | | ✓ | | | |
| **Organisation** Australian Department of Finance | | | | | | |
| **Policy Type** Social Media Guide | | | | | | |
| **Objectives** Guide for employees of the Australian Government Department when participating in online engagement | | | | | | |
| ✓ | ✓ | ✓ | ✓ | | ✓ | |
| **Organisation** The NSW Police Department | | | | | | |
| **Policy Type** Social Media Policy + Social Media Guide | | | | | | |
| **Objectives** To be used by all NSW Police Force employees when using social media | | | | | | |
| ✓ | ✓ | ✓ | ✓ | ✓ | ✓ | |
| **Organisation** Queensland Government | | | | | | |
| **Policy Type** Social Media Guide | | | | | | |
| **Objectives** The guidelines apply to all departments and covers officially established, publicly available and departmentally-managed social media accounts, but does not require the establishment of the accounts. It does not apply to use of social media on a personal or professional basis or cover use of social media for political or internal government purposes. | | | | | | |
| ✓ | ✓ | ✓ | ✓ | ✓ | ✓ | |

Table 3. Social media policies reviewed from the private sector.

| **Social Media Policies- Private Sector** |
|---|
| **Organisation** Intel |

| | Social Media Account Mgt[1] | Acceptable Use[2] | Social Media Content Mgt[3] | Employee Conduct[4] | Legal[5] | Security[6] |
|---|---|---|---|---|---|---|
| **Policy Type** Social Media Guide **Objectives** Applies to Intel employees and contractors participating in Social Media for Intel | | | | ✓ | ✓ | ✓ |
| **Organisation** Kodak **Policy Type** Social Media Guide **Objectives** Aims at Policy Developers, guiding principles | | ✓ | | ✓ | ✓ | ✓ |
| **Organisation** IBM **Policy Type** Social Media Guide **Objectives** Employees of IBM, however does not state specifically | | ✓ | | ✓ | ✓ | |
| **Organisation** Cisco **Policy Type** Social Media Policy and Social Media Guide **Objectives** Applies to all Cisco employees, vendors and contractors who are contributing or creating on social media in appropriate and effective engagement | ✓ | ✓ | ✓ | ✓ | ✓ | |
| **Organisation** Coca-Cola **Policy Type** Social Media Guide **Objectives** Applies to company employees and agency associates when using social media to help market the company and promote the brand | | ✓ | | ✓ | ✓ | |
| **Organisation** BBC - News **Policy Type** Social Media Guide **Objectives** Applies to BBC officials and employees such as reporters, presenters and correspondents relating to news coverage and online activity | ✓ | ✓ | ✓ | ✓ | ✓ | |
| **Organisation** Chartered Institute of Public Relations **Policy Type** Social Media Guide **Objectives** Designed to help UK members of CIPR navigate a rapidly evolving communications landscape | | ✓ | ✓ | ✓ | ✓ | ✓ |
| **Organisation** Ford **Policy Type** Social Media Guide **Objectives** Designed for personnel of Ford Motor Company when participating in online social media | | ✓ | | ✓ | ✓ | |
| **Organisation** Walmart **Policy Type** Social Media Guide **Objectives** Applies to customers using social media to contact Walmart, a small section applies to associates of Walmart responding on behalf of the company | ✓ | ✓ | | ✓ | ✓ | |
| **Organisation** Telstra **Policy Type** Social Media Policy **Objectives** Representing, Responsibility and Respect. Applies to all Telstra employees and contractors | ✓ | ✓ | ✓ | ✓ | ✓ | |
| **Organisation** Lifeline **Policy Type** Social Media Policy **Objectives** This policy is intended to provide employees, volunteers and supporters of Lifeline with clarity on the use of social media platforms | ✓ | ✓ | ✓ | ✓ | ✓ | |
| **Organisation** Football NSW **Policy Type** Social Media Policy **Objectives** This social media policy aims to provide some guiding principles to follow when using social media. It applies to the entire membership including players, coaches and referees. | | ✓ | ✓ | ✓ | ✓ | |

### 2.3 Technology

Traditional network security techniques such as anti-virus, firewalls and access control now combine focus with the vigilance of monitoring operations. This includes, but is not limited to, data loss prevention tools; active monitoring and analysis of security intelligence/internal auditing via penetration testers (seeking social engineering entry points). These tools will aid security professionals in becoming more accustomed to 'reading between the lines' of mined data. (Oxley, 2011). Table 4 lists the ten most applied technical controls used in Australian business, as surveyed by CERT-Australia in 2012.

Table 4. Popular technical controls used in Information Security within the enterprise.

| Technical Control | Popularity | Technical Control | Popularity |
|---|---|---|---|
| Anti-virus software | 93% | Digital Certificate | 72% |
| Firewalls | 93% | Encrypted login sessions | 68% |
| Anti-spam filters | 90% | Intrusion detection | 60% |
| Access control | 83% | Encrypted media | 50% |
| VPNs | 81% | Two factor authentication | 48% |

## 3 Key trends observed

Trends indicate that corporations are being forced to recognize emerging online threats as an overall business or strategic risk as opposed to only a technology risk. (Rudman, 2010). Overwhelmingly, all industries are discovering the need to establish trust and enhance business process transparency, due to the pervasive and instant nature of social networks. Reputation can be lost or gained rapidly through the voice of the online communities. (Almeida, 2012). The increased adoption of employees using social networking for both business benefit and personal socializing has created an immediate need for management to reassess their current security culture, making employees a primary focus. This is happening at a slower pace than these technologies are being introduced. It is evident that traditional policies and training are not effectively covering all aspects of information security. The result being, there are more cyber-attacks related to online social engineering than ever before.

Figure 1 illustrates patterns ascertained from our review of 24 enterprise social media policies and guidelines. Coverage for areas concerning ethical employee conduct and terms of acceptable use for social media usage shows prevalence. There are also high levels of inclusion relating to legal issues where coverage of applicable laws and regulations aids in online litigation protection. It is very concerning, however, that organizations are embracing social media usage without offering judicious advice relating to the security issues encountered within these technologies.

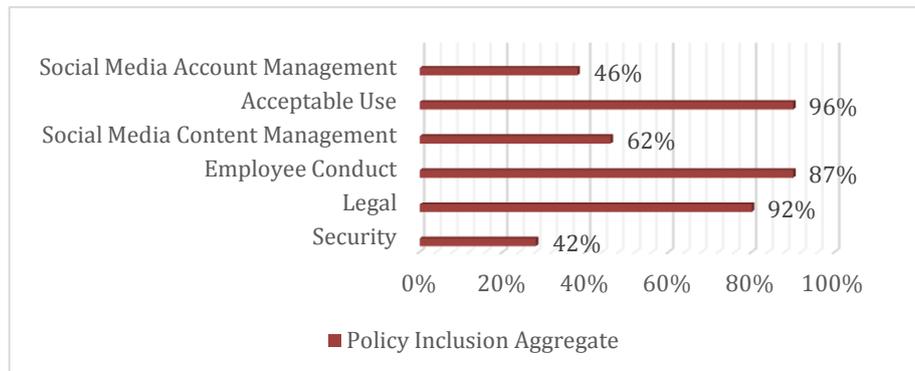

Figure 1. Comparing Social Media Policies. **Based on our tablature style review of 24 public and private sector social media policies.

From the selection of policies available, very few mentioned guidance for securing social media technologies specifically, and even fewer associated social engineering with prominent security threats. This alludes to users lacking awareness of social engineers operating over social media data, and the types of information that could be used against the employee in further attacks. They also seem to be ill-informed as to how these attacks (phishing) can be implemented and what real examples look like. What is clear from aspects of this investigation is that the problems associated with social engineering through social media are not well understood. The security obstacles continue to deepen because of the lack of clarity from management on why, when and where to apply effective countermeasures.

## 4 Conclusion

Social engineering defies traditional security efforts due to the method of attack relying on human naiveté or error. The vast amount of information now made publicly available to social engineers through online social networks is facilitating methods of attack which rely on some form of human error to enable infiltration into company networks. This investigation confirms social engineering through social media channels targeting organizational employees as one of the most challenging information security threats. There is a worrying trend to rush these technologies into the workplace without initiating effective security strategies involving social media use. We have contributed to research by addressing the gaps concerning social media policy development and the lack of advice given to employees regarding social engineering. Social engineering through social media confirms the crucial need for employees to be made aware of attack methods through a combination of policy development and employee education, alongside traditional technical countermeasures.